\documentclass{ifacconf}

\usepackage{graphicx}      % include this line if your document contains figures
\usepackage{natbib}   
\usepackage{amsmath}
\usepackage{amssymb}
\usepackage{bm}% required for bibliography
\usepackage{multirow}
\usepackage{colortbl,xcolor} 
\usepackage{enumitem}

\begin{document}
\begin{frontmatter}

\title{Data-driven control-oriented modelling for MPC-based control of urban drainage systems\thanksref{footnoteinfo}} 

\thanks[footnoteinfo]{The authors wish to acknowledge the support from the European Commission research grant of project LIFE RUBIES (LIFE20 ENV/FR/000179).
This work has
been submitted to IFAC for possible publication}

\author[IRI]{Luis Romero-Ben} 
\author[CET]{Bernat Joseph-Duran} 
\author[Aquatec]{David Sunyer}
\author[IRI]{Gabriela Cembrano} 
\author[CET]{Jordi Meseguer} 
\author[IRI]{Vicenç Puig}
\author[Canal]{Alejandro Carrasco}

%\address[UPC]{Department of Systems Eng., Automation and Industrial Informatics (ESAII),
%Universitat Politècnica de Catalunya - BarcelonaTech (UPC), Campus Sud, Building PG, Av. Diagonal 647, 08028 Barcelona, Spain (e-mail: luis.romero.ben@upc.edu).}
\address[IRI]{Institut de Robòtica i Informàtica Industrial, CSIC-UPC, Llorens i Artigas 4-6, 08028, Barcelona, Spain (e-mail: luis.romero.ben@upc.edu).}
\address[CET]{CETaqua, Water Technology Centre, Ctra. d'Esplugues 75, 08940, Barcelona, Spain.}
\address[Aquatec]{Area of Resilience and Climate Change. Aquatec S.L., Paseo de la Zona Franca 48, 08038, Barcelona, Spain.}
%\address[CS2AC]{Supervision, Safety and Automatic Control Research Center (CS2AC) - UPC, Campus de Terrassa, Rambla Sant Nebridi 22, 08222, Terrassa, Barcelona, Spain.}
\address[Canal]{Development Innovation Area. Canal de Isabel II, C/ Santa Engracia 125, 28003 Madrid, Spain.}

\begin{abstract}                % Abstract of 50--100 words
This article presents a data-driven, control-oriented modelling methodology for urban drainage systems (UDS). The proposed framework requires three main key components: input-output data from the element to be modelled, expert knowledge to define the model structure, and data-fitting techniques to obtain optimal parameters. The methodology is evaluated using a realistic benchmark from an UDS in Madrid, Spain. The results show high model accuracy and improved performance within a MPC scheme, reducing discharge and increasing treatment facilities utilization.
\end{abstract}


\end{frontmatter}
%===============================================================================

\section{Introduction}

Nowadays, global population growth and expanding urbanization %, together with challenges such as soil sealing \citep{Burghardt2006}, 
are altering the urban hydrological cycle, leading to increased combined sewer overflows (CSO) and stormwater discharge into water bodies \citep{Mcgrane2016}. The spilled water conveys pollutants like suspended solids (SS), hydrocarbons or heavy metals %\citep{Mazzarotto2021}, further degrading the water quality. %\citep{Bjorklund2018}. 
%Consequently, about 
leading to a estimated 80\% of the total volume returning to the water bodies to be polluted \citep{WWDR2017}. The most effective approach to address this challenge is the advancement in urban drainage systems (UDS) management, which transport waste and stormwater to wastewater treatment plants (WWTP) for purification before release. This management can be enhanced through efficient control strategies such as real-time control (RTC), % standing out as a particularly effective approach. RTC methods 
which iteratively monitor the system conditions and operate the actuators accordingly.

Early solutions relied on heuristic algorithms, based on the experience and intuition of network operators. The most common technique is rule-based control (RBC), which applies pre-defined rules to determine control actions under various scenarios \citep{Aulinas2011}. %, and fuzzy-logic control (FLC), integrating expert knowledge with fuzzy rules to emulate human reasoning \citep{Seggelke2013}. 
More recently, machine learning methods such as deep reinforcement learning (RL) have been exploited to develop controllers that learn from experience \citep{Mullapudi2020}. Although these approaches can be effective, the former require of extremely detailed expert knowledge and the latter can be computationally expensive during training. 

Between the development of heuristic and learning-based methods, optimization-based methods appeared as a reliable and effective solution. They use optimization algorithms and mathematical models of the UDS to minimize a certain cost function. %Such methods include linear quadratic regulators (LQR) \citep{Marinaki2003}, and model predictive control (MPC) \citep{Joseph2015}. These 
Methods such as model predictive control (MPC) \citep{Joseph2015} generally achieve excellent performance, but their applicability to real networks may be limited by the %difficulty of deriving accurate control-oriented models. This is caused by the 
complexity of real-world UDS and the associated mathematical models, as well as potential modelling errors and uncertainty. While detailed dynamic models of UDS hydrodynamics %, based on the Saint Venant equations, 
are available for urban drainage simulation, %with the aim of implementing the resulting models into model-based RTC methods, namely MPC-based schemes. 
simpler but representative models of the systems dynamics are required for optimization-based RTC, as the optimization process requires numerous computations of the outcomes of possible control actions within a control interval of a few minutes.

In this context, this article addresses the control-oriented modelling of UDS in the context of MPC-based methods. The paper presents several contributions:

\begin{itemize}
    \item \textit{A methodology to derive data-based equations that approximate the behaviour of complex network elements is presented}. Moreover, it is extended for the design of flow-setpoint curves that allow to retrieve the actuator setpoints from their associated flows.
    \item \textit{The validation of the methodology is performed on a real-world case study from Madrid}. Actual rainfall data is used to obtain results comparable to that of the real-world operation. These results are presented from a model fitting perspective, but also from a control-oriented point of view. 
\end{itemize}

\section{Methodology}

Within a MPC-based approach, the %optimization problem is defined by a set of objectives and constraints. The 
general optimal control problem can be posed by means of a state-space discrete-time model as follows: 

\begin{subequations}
\label{eq:2.2-1}
\begin{align}
% \begin{aligned}
\label{eq:2.2-1_a}
\min_{\bm{x}(k),\bm{u}(k)} \quad & J(k) \\
\label{eq:2.2-1_b}\textrm{s.t.} \quad & \bm{x}(k+1)=\mbox{\textbf{f}}\left(\bm{x}(k),\bm{u}(k),\bm{w}(k)\right)    \\
\label{eq:2.2-1_c} & \mbox{\textbf{h}}\left(\bm{x}(k),\bm{u}(k),\bm{w}(k)\right)\geq \bm{0}_{},\\
\label{eq:2.2-1_d}& \mbox{\textbf{g}}\left(\bm{x}(k),\bm{u}(k),\bm{w}(k)\right) = \bm{0}_{},
\\
\label{eq:2.2-1_e}& \bm{x_{min}}\leq \bm{x}(k)\leq \bm{x_{max}},
\\
\label{eq:2.2-1_f}& \bm{u_{min}}\leq \bm{u}(k)\leq \bm{u_{max}}
% \end{aligned}
\end{align}
\end{subequations}

\noindent where $\bm{x}(k)$\footnote{Note that $k = t, t+1, ..., t+H-1$ is the optimization time step, with $t$ being the current time instant of the event, and $H$ being the optimization horizon.} is the state vector, containing variables such as the tanks volumes, $\bm{u}(k)$ is the control input vector, containing variables representing the flows through actuators, and $\bm{w}(k)$ is the disturbance vector, related to rain runoff and intensity. The control-oriented model is set as a constraint of the problem through functions \textbf{f}, \textbf{h} and \textbf{g} (with $\bm{x_{min}}, \bm{x_{max}}, \bm{u_{min}}$ and $\bm{u_{max}}$ being the physical limits of the system elements). %, namely storage tanks and actuators)
These equations have to approximate the network behaviour while remaining simple, prioritizing linear and convex functions. %Besides improve scalability to large systems and reduce computation times, making it possible to include them within RTC implementations.  

\subsection{Simplified conceptual modelling}

A number of typical network elements can be represented using basic hydraulic models.

\subsubsection{Mass balance at junctions.} A junction node is defined as the convergence point of two or more pipes. The mass balance equation relates its inflows and outflows as follows:

\begin{equation}\label{eq:2.1.1-1}
    \sum_{i=1}^{n_{in}} q_i^{in}(k) = \sum_{j=1}^{n_{out}} q_j^{out}(k),
\end{equation}

\noindent where $q_i^{in}$ and $q_j^{out}$ are respectively the inflow from the $i$-$th$ pipe and the outflow through the $j$-$th$ pipe, and $n_{in}$ and $n_{out}$ denote the total number of inflows and outflows.

\subsubsection{Volume dynamics at the tanks} Storage tanks temporarily retain polluted water, allowing partial control of the downstream flow. Their volume evolves as follows:

\begin{subequations}
\label{eq:2.1.1-2}
\begin{align}
    \label{eq:2.1.1-2_a}
    v(k+1) = v(k) + \Delta t\left(\sum_{i=1}^{n_{in}} q_i^{in}(k) - \sum_{j=1}^{n_{out}} q_j^{out}(k)\right),\\ 
    \label{eq:2.1.1-2_b}
    0 \leq q_i^{in}(k) \leq q^{in,max}, \\
    \label{eq:2.1.1-2_c}
    0 \leq q_j^{out}(k) \leq q^{out,max},
\end{align}
\end{subequations}
%    \label{eq:2.1.1-2_d}
    %0 \leq v(k) \leq v^{max},

\noindent where $v$ denotes the tank volume, $q_i^{in}$ and $q_i^{out}$ represent, in this case, the inflows and outflows of the tank (limited by $q^{in,max}$ and $q^{out,max}$), which are shaped through actuators, and $\Delta t$ is the sampling time.

\subsubsection{Combined sewer overflow} CSOs release excess water into natural bodies when the physical or operational capacity of network components is surpassed:

\begin{itemize}
    \item Storage tanks include an overflow mechanism to prevent volume from exceeding the maximum capacity $v_{max}$. This outfall can be modelled as:
    \begin{equation}\label{eq:2.1.1-4}
        q^{CSO}(k) = \frac{1}{\Delta t}\mbox{max}\left(0,v(k+1)-v^{max}\right),
    \end{equation}

    \noindent where $v(k+1)$ is defined through \eqref{eq:2.1.1-2_a} and $v^{max}$ is the physical limit of the tank storage capacity. % by means of the current tank inflows and outflows.
    \item WWTPs and CSO points can overflow when the incoming flow exceeds their capacity. This overflow can be defined as:

    \begin{subequations}\label{eq:2.1.1-5}
    \begin{align}
        \label{eq:2.1.1-5_a}q^{out}(k) = \mbox{min}\left(q^{in}(k),q^{max}\right),\\
        \label{eq:2.1.1-5_b}q^{CSO}(k) = q^{in}(k) - q^{out}(k),
    \end{align}
    \end{subequations}
    
    \noindent where $q^{in}$ and $q^{out}$ are the flow upstream and downstream to the CSO point, respectively. Besides, $q^{max}$ is the maximum capacity of the network element.
\end{itemize}

\subsection{Data-driven modelling}

However, some elements in an UDS may have complex dynamics which cannot be captured by simplified conceptual models, such as active components (e.g. pumps and valves) and standard elements with a complex behaviour under demanding conditions, like collectors running full during intense rainfall. In these cases, data-based models provide an effective alternative. The derivation of these models requires three key sources of information: (i) large volumes of input-output data from the component to model, covering a wide range of operating conditions; (ii) expert knowledge to guide the modelling process and define parametrized expressions consistent with the element's behaviour; (iii) data-fitting algorithms that identify patterns from the data, estimating the optimal parameter values.

Based on these requisites, a framework for deriving data-driven models can be defined. The preliminary step involves analysing the network to identify which components require data-driven modelling. Then, the following steps are applied to each selected element:

\begin{enumerate}
    \item \textbf{The necessary input-output data are gathered}, considering multiple rain scenarios of different characteristics. Although measurements can be used if sensors are installed, input-output synthetic data is often generated through a high-fidelity detailed network hydrodynamics model under realistic conditions.
    \item \textbf{The input-output data are graphically represented}, assigning each variable to an axis. In this way, experts can analyse the element's behaviour, and the relation between inflows and outflows. For elements with multiple inputs/outputs, the analysis can be performed iteratively, considering different input/output pairs and studying the effect of each input over each output.
    \item \textbf{The insights from the data are used to select a mathematical expression} that can represent the component's behaviour. This includes deciding the type of function to be used and its parametrization: 
    \begin{itemize}
        \item Certain mathematical functions are not desirable for optimization-oriented models because they may pose problems for convergence. Preferred expressions include (in decreasing order): (a) linear functions, (b) polynomial functions, (c) logarithmic functions, and (d) exponential functions.
        \item Moreover, a trade-off must be considered in the number of parameters, as excessive parametrization increases complexity and even may degrade the obtained solution.
    \end{itemize}
    \item \textbf{A data-fitting algorithm is applied} to get the optimal set of parameter values. Various types of algorithms are described in Table \ref{tb:algorithms}, namely linear (LLS) %least squares%(solved analytically, or through QR or singular value decomposition)
    and non-linear (NLLS) least squares \citep{Hansen2013}, %(e.g. Gauss-Newton, Levenberg-Marquardt), 
    non-parametric \citep{Eubank1999}, %(e.g. splines) 
    and heuristic %(e.g. genetic algorithms) 
    fitting \citep{Gulsen1995}. 
    \item \textbf{The obtained data-based equation is evaluated} over calibration and test rain scenario data, comparing real and predicted output values.
\end{enumerate}

\begin{table*}[htb]
\begin{center}
\caption{Data-fitting algorithms: (i) LLS, (ii) NLLS, (iii) non-parametric and (iv) heuristic. }\label{tb:algorithms}
\begin{tabular}{ccll}
& \textbf{Use} & \textbf{Advantages} & \textbf{Limitations}  \\\hline
\arrayrulecolor{gray!60}
(i) & Linear parametrized models & Extremely efficient, with closed-form solution& Non-linear systems, outliers and noise\\\hline
 
 (ii) & Non-linear parametrized models & Robust, works for numerous classes of models  & May be costly or non-optimal\\\hline

(iii)  & Unknown function form & Flexible, minimal assumptions on expressions & Risk of overfitting, loss of expert insight \\\hline

(iv) & Non-smooth objective & May avoid local minima, derivative-free & Computationally costly, time-consuming\\

\arrayrulecolor{black}\hline
\end{tabular}
\end{center}
\end{table*}

% \begin{figure}
% \begin{center}
% \includegraphics[width=8.4cm]{diagrama_databased.pdf}    % The printed column width is 8.4 cm.
% \caption{Operational flow of the data-driven modelling methodology} 
% \label{fig:data-based}
% \end{center}
% \end{figure}

\subsubsection{Flow-setpoint conversion}

%Among the network components requiring data-based modelling, actuators represent a special case. 
In UDS, actuators are mainly associated with tank operations, namely \textit{filling}, i.e., storing polluted water during rainfall; and \textit{emptying}, i.e.,  releasing it when treatment facilities have sufficient capacity. These operations are usually executed through pumps (with adjustable curves) or orifices (with controllable aperture level). The tank dynamics, described in \eqref{eq:2.1.1-2}, depend on the inflows and outflows, which are assigned certain values within the MPC prediction to optimize the tank operation according to the cost function. Thus, these predicted flows must be converted into actuator setpoints for their application in the subsequent control step, leading approximately to the optimized flow targets. A similar procedure to that described earlier can be applied to derive the flow–setpoint conversion function, but various updates are required:

\begin{enumerate}[label=\Roman*.]
    \item The relationship between actuator inflows and outflows may be affected by additional factors modifying its operation, such as saturations during intense rainfalls. Thus, other network variables may have to be incorporated to the input-output datasets.
    \item When generating synthetic data, the full range of setpoint values must be tested to accurately reconstruct the actuator's behaviour. This is achieved by defining a discrete grid of setpoint values (e.g. orifice openings with fixed increments). To reduce computational cost, separate datasets can be obtained for each actuator, varying only the studied actuator while controlling others through expert-based rules.
    \item The generated data must be grouped into batches, depending on the specific actuator value, allowing the analysis of each specific actuator configuration independently. This helps to understand the shape of the function that best models the actuator behaviour depending on the setpoint value. Such analysis would require to compare the flow related to the control action against all the other involved variables.
    %\item Finally, data-fitting is performed by supplying the gathered and pre-processed data to the selected data-fitting algorithm. 
\end{enumerate}

We must remark that a conversion function is derived for each discrete setpoint value of each actuator. During application, the input variables for the conversion function are obtained from the optimized results (at the next prediction step), and the conversion functions for all the setpoints are applied over them. The obtained values are then compared against the optimized value for the actuator flow variable (normally its outflow). The final setpoint value is selected depending on a pre-defined criterion, such as choosing the discrete setpoint value that provides the closest conversion function value to the optimized value, or even interpolating a setpoint value between the lower and upper discrete setpoint values. 

\section{Case study}

The proposed methodology is assessed using a case study implemented on a fraction of Madrid's city UDS. Specifically, we use a reduced part of the network, mainly composed of two interceptors, parallel to the Manzanares river by its left and right margins (i.e., LM and RM), where the filling and emptying of two tanks are regulated using actuators. The main properties of this pilot are summarised in Table \ref{tb:network}. A detailed UDS model was implemented using Storm Water Management Model (SWMM) \citep{Rossman2015}. This model includes the section considered as pilot, but also the rest of the UDS, which generates the inflows to the actual pilot. Note that actuators outside the pilot are controlled by pre-defined rules.

\begin{table}[htb]
\begin{center}
\caption{Characteristics of the case study UDS}\label{tb:network}
\begin{tabular}{ccl}
\textbf{Element} & \textbf{Nº} & \textbf{Characteristics}  \\\hline
\arrayrulecolor{gray!60}

Tank & 2 & Butarque [RM, $3\cdot10^5\: m^3$ max. capacity   \\
&  &  (MC)], Abroñigales [LM, $2\cdot10^5\: m^3$ MC]\\\hline

WWTP & 3 & Butarque (RM) and Sur (LM), total\\
&  &  $4.75\cdot10^5\: m^3$/day MC; La Gavia (LM)\\\hline

Inlets & 7 & 4 in RM and 3 in LM \\\hline

CSOs & 6 & 3 in RM and 3 in LM \\\hline

\arrayrulecolor{black}\hline
\end{tabular}
\end{center}
\end{table}

% Actuators & 4 & Butarque filling (pump) and emptying    \\
% &  & (orifice), Abroñigales filling (orifice)\\
% &  & and emptying (orifice) \\\hline

\section{Results and discussion}

\begin{figure}
\begin{center}
\includegraphics[width=8.4cm]{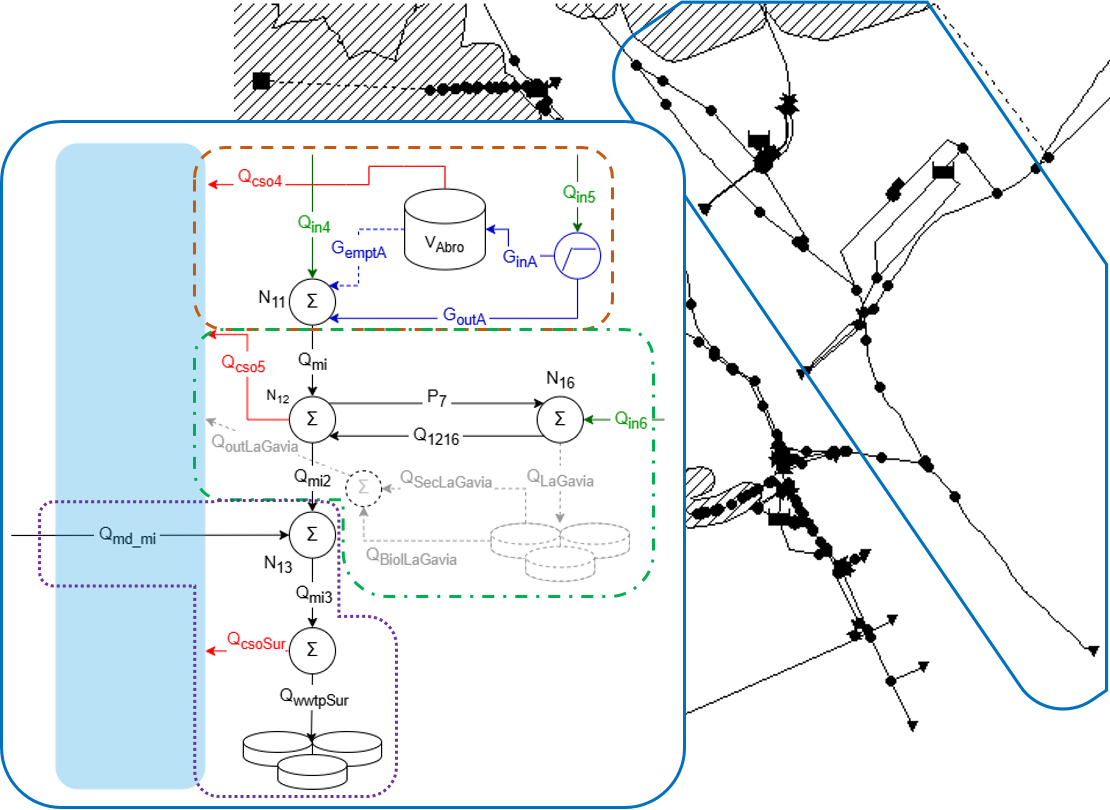}    % The printed column width is 8.4 cm.
\caption{Schematic representation of the simplified model of the left bank, depicted over the detailed model.} 
\label{fig:network}
\end{center}
\end{figure}

The evaluation of the modelling framework is carried out from data-fitting and control-oriented perspectives. %through model calibration analysis and a subsequent evaluation of a MPC strategy for UDS using the obtained simplified model. 
Although a simplified model was obtained for the complete pilot with its two margins, for the sake of simplicity, we focus here on LM, whose graphical representation is shown over the detailed model in Fig.~\ref{fig:network}. 

\subsection{Modelling}

The LM model can be divided into three main areas: upstream LM and Abroñigales tank (dashed enclosure in Fig.~\ref{fig:network}), La Gavia WWTP area (dashed-dotted enclosure), and RM - LM connection and Sur WWTP area (dotted enclosure). The model equations %, divided into these three areas, 
are presented in Table \ref{tb:equations}, using the variable names in Fig.~\ref{fig:network}. % are used to denote LM model variables. 
%Please refer to the figure to visualize the relationships among variables and their role within the LM model.

\begin{table*}[htb]
\begin{center}
\caption{Summary of the simplified model of LM, indicating each equation's description, type (SC = simplified conceptual, DB = data-based), and formula.}\label{tb:equations}
\begin{tabular}{lcr}
\textbf{Description} & \textbf{Type} & \textbf{Equation} \\\hline
\multicolumn{3}{c}{\textbf{\textit{Upstream LM and Abroñigales tank}}} \\\hline
\arrayrulecolor{gray!60}
Mass balance in $N_{11}$ & SC \eqref{eq:2.1.1-1} & $\displaystyle
\refstepcounter{equation}\label{eq:3.1-1}
Q_{mi}(t) = Q_{in4}(t) + G_{emptA}(t) + G_{outA}(t) \quad \text{(\theequation)}$ \\\hline

Tank inflow and bypass from $Q_{in5}$ & SC \eqref{eq:2.1.1-1} & $\displaystyle
\refstepcounter{equation}\label{eq:3.1-2}
G_{inA}(t) = Q_{in5}(t) - G_{outA}(t) \quad \text{(\theequation)}$
 \\\hline
 
Abroñigales tank volume dynamics & SC \eqref{eq:2.1.1-2_a} & $\displaystyle
\refstepcounter{equation}\label{eq:3.1-3}
V_{Abro}(t+1) = V_{Abro}(t) + \Delta t\left(G_{inA}(t) - G_{emptA}(t) - Q_{CSO4}(t)\right)
\quad \text{(\theequation)}$\\\hline

Abroñigales tank outflow limit & SC \eqref{eq:2.1.1-2_c} & $\displaystyle
\refstepcounter{equation}\label{eq:3.1-4}
0 \leq G_{emptA}(t) \leq V_{Abro}(t)/\Delta t
\quad \text{(\theequation)}$\\\hline

Abroñigales tank associated CSO & SC \eqref{eq:2.1.1-4} & $\displaystyle
\refstepcounter{equation}\label{eq:3.1-5}
Q_{CSO4}(t) = \max(0,\,
V_{Abro}(t+1)
- V_{Abro}^{max})/\Delta t
\quad \text{(\theequation)}$
\\\arrayrulecolor{black}\hline

\multicolumn{3}{c}{\textbf{\textit{La Gavia WWTP area}}} \\\hline
\arrayrulecolor{gray!60}

Computation of $Q_{1216}$ from $Q_{in6}$ & DB & $\displaystyle
\refstepcounter{equation}\label{eq:3.1-6}
Q_{1216}(t) = \max(0,\; 0.003Q_{in6}^2(t) + 0.921Q_{in6}(t) - 0.538)
\quad \text{(\theequation)}$
 \\\hline

Mass balance in $N_{16}$ (max. outflow) & SC (\ref{eq:2.1.1-1},\ref{eq:2.1.1-5_a}) & $\displaystyle
\refstepcounter{equation}\label{eq:3.1-7}
Q_{LaGavia}(t) = \min\left(1.5,\; P_7(t) + \left(Q_{in6}(t) - Q_{1216}(t)\right)\right)
\quad \text{(\theequation)}$\\\hline

Biological treatment in La Gavia & SC \eqref{eq:2.1.1-5_a} & $\displaystyle
\refstepcounter{equation}\label{eq:3.1-8}
Q_{BiolLaGavia}(t) = \min\left(1.25,\; Q_{LaGavia}(t)\right)
\quad \text{(\theequation)}$
\\\hline

Secondary treatment in La Gavia & SC \eqref{eq:2.1.1-1} & $\displaystyle
\refstepcounter{equation}\label{eq:3.1-9}
Q_{SecLaGavia}(t) = Q_{LaGavia}(t) - Q_{BiolLaGavia}(t)
\quad \text{(\theequation)}$
\\\hline

Flow to CSO from La Gavia & SC \eqref{eq:2.1.1-1} & $\displaystyle
\refstepcounter{equation}\label{eq:3.1-10}
Q_{outLaGavia}(t) = Q_{SecLaGavia}(t) - Q_{BiolLaGavia}(t)
\quad \text{(\theequation)}$
\\\hline

Downstream from $N_{12}$ & DB & $\displaystyle
\refstepcounter{equation}\label{eq:3.1-11}
Q_{mi2}(t) = -0.003Q_{1216}^2(t) - 0.041Q_{mi}^2(t) + 0.167Q_{1216}(t) + 0.874Q_{mi}(t)
\quad \text{(\theequation)}$
\\\hline

CSO discharge from $N_{12}$ & DB & $\displaystyle
\refstepcounter{equation}\label{eq:3.1-12}
Q_{CSO5.1}(t) = 0.006Q_{1216}^2(t) + 0.042Q_{mi}^2(t) + 0.707Q_{1216}(t) + 0.155Q_{mi}(t)
\quad \text{(\theequation)}$
\\\hline

Computation of $Q_{CSO5}$ & SC \eqref{eq:2.1.1-1} & $\displaystyle
\refstepcounter{equation}\label{eq:3.1-13}
Q_{CSO5}(t) = Q_{CSO5.1}(t) + Q_{outLaGavia}(t)
\quad \text{(\theequation)}$
\\\hline

$N_{12}$ water level from flows & DB & $\displaystyle
\refstepcounter{equation}\label{eq:3.1-14}
L_{7}(t) = -0.254Q_{mi2}(t) - 0.6Q_{CSO5}(t) + 0.583Q_{1216}(t) + 0.77Q_{mi}(t) + 0.868
\quad \text{(\theequation)}$
\\\hline

Pump $P_7$ level-to-flow curve & DB & $\displaystyle
\refstepcounter{equation}\label{eq:3.1-15}
P_{7}(t) = 0.455/(0.883 + e^{-65.998L_{7}(t) + 77.339}) + 0.132
\quad \text{(\theequation)}$
\\\arrayrulecolor{black}\hline

\multicolumn{3}{c}{\textbf{\textit{RM-LM connection and Sur WWTP area}}} \\\hline
\arrayrulecolor{gray!60}

LM downstream flow & SC \eqref{eq:2.1.1-1} & $\displaystyle
\refstepcounter{equation}\label{eq:3.1-16}
Q_{mi3}(t) = Q_{mi2}(t) + Q_{md\_mi}(t)
\quad \text{(\theequation)}$
 \\\hline

Flow entering Sur WWTP & SC \eqref{eq:2.1.1-5_a} & $\displaystyle
\refstepcounter{equation}\label{eq:3.1-17}
Q_{WWTPSur}(t) = \min(6,\; Q_{mi3}(t))
\quad \text{(\theequation)}$
 \\\hline

CSO from Sur WWTP & SC \eqref{eq:2.1.1-5_b} & $\displaystyle
\refstepcounter{equation}\label{eq:3.1-18}
Q_{CSOSur}(t) = Q_{mi3}(t) - Q_{WWTPSur}(t)
\quad \text{(\theequation)}$
 \\\hline

\arrayrulecolor{black}\hline
\end{tabular}
\end{center}
\end{table*}

\subsubsection{Upstream LM and Abroñigales tank} This section contains two of the water inlets within LM. %, arriving from different parts of the UDS system. 
Firstly, the upstream flow to LM arrives through $Q_{in4}$, reaching virtual node\footnote{Virtual nodes are simplifications of actual connections in the detailed model, used to represent the flow balance.} $N_{11}$, % , which encompasses the two trunks in which the upstream flow gets divided in the detailed model. 
which also receives water from the Abroñigales tank outflow ($G_{emptA}$) and bypass ($G_{outA}$), yielding $Q_{mi}$ (eq. \eqref{eq:3.1-1}). Secondly, the water coming through $Q_{in5}$ reaches the diversion point between Abroñigales tank inflow ($G_{inA}$) and bypass, where this division is controlled using an actuator (orifice) (eq. \eqref{eq:3.1-2}). Abroñigales tank volume evolves depending on its inflows and outflows, physical/operational limits and related CSO (eqs. (\ref{eq:3.1-3}-\ref{eq:3.1-5})).

\subsubsection{La Gavia WWTP area} The last water inlet is $Q_{in6}$, whose flow is passively divided into water bypassed downstream ($Q_{1216}$) and water conveyed to La Gavia WWTP area. The relationship between $Q_{in6}$ and $Q_{1216}$ was modelled using data-fitting due to its complexity, which cannot be properly represented using mass conservation only %Besides, the division of $Q_{in6}$ is not defined by an active element. Thus, data-based fitting was used to model the relationship between $Q_{in6}$ and $Q_{1216}$
(eq. \eqref{eq:3.1-6}). Then, we apply mass balance in $N_{16}$, regarding the physical limit to La Gavia inflow (eq. \eqref{eq:3.1-7}). This flow is divided into the water receiving biological or secondary treatments, as well as the WWTP outfall (eqs. (\ref{eq:3.1-8}-\ref{eq:3.1-10})). Next, two data-based equations are generated from $N_{12}$, using $Q_{1216}$ and the flow upstream ($Q_{mi}$) to yield the flow downstream ($Q_{mi2}$) and the remaining portion of $Q_{CSO5}$, added to La Gavia outfall to yield the complete CSO flow (eqs. (\ref{eq:3.1-11}-\ref{eq:3.1-13})). Finally, pump $P_7$ is modelled through a pair of data-based equations: the first converts the flows related to $N_{12}$ into a ``virtual" water level (eq. \eqref{eq:3.1-14}), and the second mimics the actual pumping curve, which converts node level into pumped flow (eq. \eqref{eq:3.1-15}).

\subsubsection{RM-LM connection and Sur WWTP area} This final part of LM is defined by the operation of Sur WWTP. The water conveyed through $Q_{mi2}$ is added to the water coming from RM (eq. \eqref{eq:3.1-16}) and then Sur WWTP's inflow is defined by its capacity, with the rest of incoming water being released as CSO (eqs. (\ref{eq:3.1-17}-\ref{eq:3.1-18})).

\subsubsection{Discussion} From a modelling perspective, we can evaluate the goodness of fitting of the data-based equations, namely (\ref{eq:3.1-6}, \ref{eq:3.1-11}, \ref{eq:3.1-12}, \ref{eq:3.1-14},  \ref{eq:3.1-15}) in Table \ref{tb:equations}. To calibrate them, outflow and inflow/s data are gathered from available rain events, which are shown in Table \ref{tb:rain}. Besides, all potential setpoint combinations must be considered. We define 10\% opening step in each actuator, leading to a total of 100 scenarios per rain, considering the two available actuators (i.e., orifices associated to $G_{outA}$ and $G_{emptA}$).

\begin{table}[htb]
\begin{center}
\caption{Precipitation $P$ (mm), maximum intensity $I_{max}$ (mm/h), date and duration of the calibration rains scenarios} \label{tb:rain}
\begin{tabular}{cccc}
$\bm{P}$ & $\bm{I_{max}}$ & \textbf{Date} & \textbf{Duration}  \\\hline
\arrayrulecolor{gray!60}

17.0 & 31.2 & 2016-10-22 & 10h 10min \\\hline
35.8 & 76.8 & 2017-01-09 & 17h 20min \\\hline
23.6 & 50.4 & 2017-02-07 & 5h 35 min \\\hline
15.4 & 33.6 & 2017-12-12 & 3h 50 min \\\hline
21.3 & 36.0 & 2018-03-04 & 13h 45 min \\\hline
19.4 & 60.0 & 2018-03-10 & 1h 5 min \\

\arrayrulecolor{black}\hline
\end{tabular}
\end{center}
\end{table}

Data-fitting performance metrics are presented in Table \ref{tb:model_performance}, namely root-mean squared error ($RMSE$), 
mean absolute error ($MAE$) and coefficient of determination ($R^2$). The first two provide error measures in flow units ($m^3/s$), with the former being more affected by outliers, showing if there are occasional large errors; and the latter being more robust, increasing interpretability. $R^2$ measures how well the model explains the variance in data. The mean, standard deviation and maximum value of the true outcome of each expression are also given in order to enrich the analysis. 

\begin{table}[htb]
\begin{center}
\caption{Summary of model performance metrics.} \label{tb:model_performance}
\begin{tabular}{cccccc}
\textbf{Eq.} & $\bm{\overline{y}\pm\sigma(\bm{y})}$ & $\bm{\max(\bm{y})}$ & $\bm{RMSE}$ & $\bm{MAE}$ & $\bm{R^2}$  \\\hline
\arrayrulecolor{gray!60}

\eqref{eq:3.1-6} & $3.28\pm5.01$ & 21.26 & 0.44 & 0.21 & 0.99 \\\hline

\eqref{eq:3.1-11} & $4.18\pm1.41$ & 7.54 & 0.39 & 0.32 & 0.92 \\\hline

\eqref{eq:3.1-12} & $6.86\pm6.55$ & 27.83 & 0.95 & 0.51 & 0.97 \\\hline

\eqref{eq:3.1-14} & $3.61\pm1.19$ & 5.61 & 0.09 & 0.06 & 0.99 \\\hline

\eqref{eq:3.1-15} & $0.64\pm0.05$ & 0.65 & $9\cdot10^{-3}$ & $4\cdot10^{-3}$ & 0.97 \\

\arrayrulecolor{black}\hline
\end{tabular}
\end{center}
\end{table}

Analysing Table \ref{tb:model_performance}, we can appreciate how each expression excellently fits the associated input-output data. All the data-based models have $R^2>0.9$, with small errors in comparison to the mean values. Regarding the resulting values, \eqref{eq:3.1-11} would be the expression with the most limited performance, although its metrics remain satisfactory. \eqref{eq:3.1-6} and \eqref{eq:3.1-14} stand out due to their notable capabilities.

\subsection{Control}

The data-fitting results can be complemented with the MPC control performance under a realistic rain scenario. Besides, this scenario can be used to evaluate the flow-setpoint conversion functions for the two actuators included within the case study, defined by the filling and emptying operations of Abroñigales.

\subsubsection{Filling operation} The inflow to the tank is defined in \eqref{eq:3.1-2}. %, with the optimizer computing the optimal value of the involved variables to minimize the cost function. In this case, t
The actuator setpoint is the opening level of the associated orifice, with $0\%$ implying that the bypass $G_{outA}$ is closed. % , and $100\%$ denoting a fully open bypass. 
Thus, a function converting $Q_{in5}$ into $G_{outA}$ must be defined, namely: %. After analysing the input-output relation, the following data-based model was derived:

\begin{equation}\label{eq:4.2-1}
    G_{outA}(t) = \begin{cases}
    m_i\left(Q_{in5}(t)-x_i\right)+y_i,& Q_{in5}(t)<p_i \\
    \mbox{f}_2\left(Q_{in5},a_i,b_i,c_i,d_i,e_i\right),& \mbox{otherwise}
    \end{cases}
\end{equation}

\noindent where $\mbox{f}_2\left(Q_{in5},a_i,b_i,c_i,d_i,e_i\right)=a_iQ_{in5}^2(t)+b_iQ_{in5}(t)+c_i+d_i\left(\mbox{ln}(e_iQ_{in5}(t))\right)$, with parameters depending on $i$ because they belong to a different potential setpoint value on the defined grid (with a step of $10\%$). The equation represents two distinct modes: (i) a linear expression, describing the behaviour when the bypass can accommodate the water coming from $Q_{in5}$; and (ii) a saturation expression, representing system overload at the diversion node (with the rest of water entering the tank). The parameter values and discontinuity points of are given in Table \ref{tb:parametros_GoutA}.%\textbf{In practice, due to the characteristics of real rainfalls and the pilot (and detailed model) dynamics, the linear model is almost always active: in dry weather, the reduced flow in $Q_{in5}$ does not normally surpass the discontinuity point $p_i$, and during rain events, the controller tries to fill the tank as much as it can, leading to low (normally null) setpoints to $G_{outA}$ despite the saturation model would be active due to $Q_{in5}$. Therefore, for clarity, only the parameter values of the linear expression are given in Table \ref{tb:parametros_GoutA}. }

% \begin{table}[htb]
% \begin{center}
% \caption{Parameter values for \eqref{eq:4.2-1}.} \label{tb:parametros_GoutA}
% \begin{tabular}{cccccc}
% \hline
% $i$ & \% & $m_i$ & $x_i$ & $y_i$ & $p_i$ \\\hline
% \arrayrulecolor{gray!60}
% 0  & 0   & 0.00 & 0.26 & 0.10 & 0.00 \\\hline
% 1  & 10  & 0.59 & 0.26 & 0.10 & 0.90 \\\hline
% 2  & 20  & 0.93 & 0.26 & 0.17 & 1.10 \\\hline
% 3  & 30  & 0.94 & 0.26 & 0.21 & 1.50 \\\hline
% 4  & 40  & 1.04 & 0.26 & 0.24 & 1.80 \\\hline
% 5  & 50  & 1.05 & 0.26 & 0.26 & 2.20 \\\hline
% 6  & 60  & 1.10 & 0.26 & 0.26 & 2.50 \\\hline
% 7  & 70  & 1.10 & 0.26 & 0.26 & 2.90 \\\hline
% 8  & 80  & 1.06 & 0.26 & 0.26 & 3.40 \\\hline
% 9  & 90  & 1.09 & 0.26 & 0.26 & 3.65 \\\hline
% 10 & 100 & 1.08 & 0.26 & 0.26 & 4.00 \\
% \arrayrulecolor{black}\hline
% \end{tabular}
% \end{center}
% \end{table}

\subsubsection{Emptying operation} The outflow of the tank depends on factors such as the tank water level ($D_{Abro}$) and other flows related to \eqref{eq:3.1-1}, namely $Q_{in4}$ and $G_{outA}$. In this case, the actuator setpoint is the opening of the associated orifice, which is directly related to $G_{emptA}$, and the conversion function relates the mentioned factors with this flow: %. An analysis of the input-output data led to the next expression:

\begin{equation}\label{eq:4.2-2}
\begin{split}
    G_{emptA}(t) = f_iD_{Abro}^2(t)+g_iD_{Abro}(t)+\\h_iG_{outA}(t)+r_iQ_{in4}(t)+s_i,
\end{split}
\end{equation}

\noindent where the parameter values are showed in Table \ref{tb:parametros_GoutA}.

\begin{table*}[htb]
\begin{center}
\caption{Parameter values and discontinuity points for \eqref{eq:4.2-1} and \eqref{eq:4.2-2}.}
\label{tb:parametros_GoutA}
\begin{tabular}{c|ccccccccccc}

\textbf{i} & \textbf{0} & \textbf{1} & \textbf{2} & \textbf{3} &\textbf{4} & \textbf{5} & \textbf{6} & \textbf{7} & \textbf{8} &\textbf{ 9} & \textbf{10} \\\hline
\%     & 0 & 10 & 20 & 30 & 40 & 50 & 60 & 70 & 80 & 90 & 100 \\\hline\arrayrulecolor{gray!60}
$m_i$  & 0 & 0.59 & 0.93 & 0.94 & 1.04 & 1.05 & 1.10 & 1.10 & 1.06 & 1.09 & 1.08 \\\hline
$x_i$  & 0.26 & 0.26 & 0.26 & 0.26 & 0.26 & 0.26 & 0.26 & 0.26 & 0.26 & 0.26 & 0.26 \\\hline
$y_i$  & 0.10 & 0.10 & 0.17 & 0.21 & 0.24 & 0.26 & 0.26 & 0.26 & 0.26 & 0.26 & 0.26 \\\hline
$p_i$  & 0 & 0.90 & 1.10 & 1.50 & 1.80 & 2.20 & 2.50 & 2.90 & 3.40 & 3.65 & 4.00 \\\hline
$a_i$  & 0 & $5\cdot10^{-6}$ & $3\cdot10^{-5}$ & $-2\cdot10^{-5}$ & $-9\cdot10^{-5}$ & $-2\cdot10^{-4}$ & $-9\cdot10^{-5}$ & $-2\cdot10^{-4}$ & $-6\cdot10^{-4}$ & $-3\cdot10^{-4}$ & $8\cdot10^{-6}$ \\\hline
$b_i$  & 0 & $6\cdot10^{-4}$ & $6\cdot10^{-4}$ & $4\cdot10^{-3}$ & $8\cdot10^{-3}$ & $1\cdot10^{-2}$ & $9\cdot10^{-3}$ & $2\cdot10^{-2}$ & $4\cdot10^{-2}$ & $2\cdot10^{-2}$ & $-7\cdot10^{-3}$ \\\hline
$c_i$  & 0 & 0.40 & 0.84 & 1.32 & 1.86 & 2.18 & 6.10 & 2.86 & 3.13 & 11.20 & 21.19 \\\hline
$d_i$  & 0 & 0.01 & 0.02 & 0.04 & 0.03 & 0.03 & 0.08 & 0.05 & $-8\cdot10^{-3}$ & 0.20 & 0.42 \\\hline
$e_i$  & 0.3 & $5.3\cdot10^{3}$ & $1.5\cdot10^{2}$ & 1.80 & 0.18 & 9.50 & 0 & 64.1 & 0 & 0 & 0 \\ \arrayrulecolor{black}\hline\arrayrulecolor{gray!60}

$f_i$  & 0 & -0.00 & -0.01 & -0.02 & -0.03 & -0.03 & -0.04 & -0.05 & -0.06 & -0.04 & -0.05 \\\hline
$g_i$  & 0 & 0.12 & 0.30 & 0.47 & 0.62 & 0.78 & 0.93 & 1.09 & 1.25 & 1.11 & 1.29 \\\hline
$h_i$  & 0 & -0.02 & -0.04 & -0.07 & -0.10 & -0.12 & -0.15 & -0.18 & -0.20 & -0.23 & -0.27 \\\hline
$r_i$  & 0 & -0.02 & -0.03 & -0.05 & -0.07 & -0.10 & -0.13 & -0.17 & -0.21 & -0.24 & -0.28 \\\hline
$s_i$  & 0 & 0.24 & 0.29 & 0.36 & 0.46 & 0.56 & 0.66 & 0.78 & 0.88 & 1.66 & 1.71 \\\hline
\arrayrulecolor{black}\hline
\end{tabular}
\end{center}
\end{table*}

\subsubsection{Discussion} To evaluate the goodness of conversion functions, we have implemented them within a close-loop framework, testing the complete control scheme and the derived model (including the RM model). The role of the virtual reality is played by the SWMM detailed model, whereas the MPC problem is solved using General Algebraic 
Modelling System or GAMS \citep{GAMS2024}, which is a high-level modelling software that allows to solve linear, non-linear, and even mixed-integer optimization problems. In this case, the control-oriented model is implemented as the constraints in \eqref{eq:2.1.1-2}, and the cost function $J$ has three sub-objectives: $J_{CSO}$, which seeks to minimize the CSO volume; $J_{WWTP}$, which tries to maximize WWTP utilization; and $J_{smooth}$, which seeks to minimize changes in the actuator setpoints.

\begin{figure}
\begin{center}
\includegraphics[width=8.4cm]{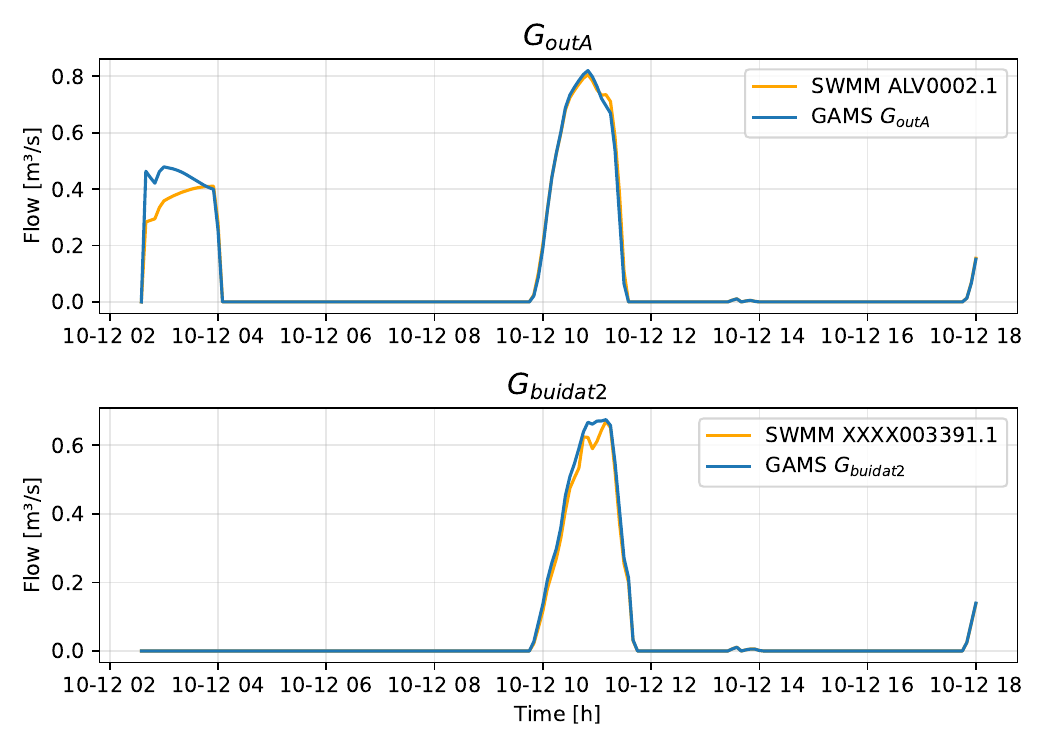}    % The printed column width is 8.4 cm.
\caption{Evolution of the actuator-related flows $G_{outA}$ and $G_{emptA}$ on the 12/10/2024 event.} 
\label{fig:conv_func}
\end{center}
\end{figure}

The methodology is tested on a rain scenario occurred on 12/10/2024, with a duration of 11 hours and 21.9 mm of precipitation. The conversion functions are assessed by comparing predicted and actual actuator flows. Note that, during application, the input variables for each function are extracted from the optimization results (at $k=t+1$), and the conversion functions for all the potential setpoints are applied over them. The obtained values are compared against $G_{outA}$ and $G_{emptA}$, and the setpoints whose conversion function lead to the closest outcomes to the optimized value are selected, interpolating the final setpoint value between the lower and upper discrete setpoint values. Thus, in Fig.~\ref{fig:conv_func}, we compare the MPC predicted flow for $G_{outA}$ and $G_{emptA}$ with the actual flow value at the associated SWMM link at the next time instant (when the final setpoint is actually applied). 
Both SWMM and GAMS lines are extremely similar, denoting the excellent conversion capability of \eqref{eq:4.2-1} and \eqref{eq:4.2-2}. The $R^2$ metric %of Table \ref{tb:model_performance} 
in the estimation is 0.98 for $G_{outA}$ and a 0.99 for $G_{emptA}$, showing their outstanding performance. 

Finally, to illustrate the advantages of deriving accurate control-oriented model and conversion functions, the overall controller performance in the considered rain event is assessed by comparing the CSO discharges and WWTP inflows for the elements in LM. These values are presented in Table \ref{tb:control_performance} for the MPC method, as well as a RBC system. The results show how the MPC controller is able of notably reducing the CSO spillage in LM, specifically by a 19.1\%; while increasing the WWTP inflow by a 7.9\%. %\footnote{Note that there are other CSO and WWTP in the complete pilot, present in RM. However, we only analyse the ones in LM to illustrate the effect of the presented modelling.}.

\begin{table}[htb]
\begin{center}
\caption{Summary of key performance indicators for RBC and MPC.} \label{tb:control_performance}
\begin{tabular}{c|cc}
 & \textbf{RBC} ($10^3 m^3$) & \textbf{MPC} ($10^3 m^3$) \\\hline
\arrayrulecolor{gray!60}

$Q_{CSO4}$ & 0 & 0 \\\hline
$Q_{CSO5}$ & 313.09 & 238.19 \\\hline
$Q_{CSOSur}$ & 74.06 & 74.96 \\
\arrayrulecolor{black}\hline
\arrayrulecolor{gray!60}
$Q_{WWTPSur}$ & 240.45 & 259.43 \\
\arrayrulecolor{black}\hline
\end{tabular}
\end{center}
\end{table}

\section{Conclusion}

This article presents a control-oriented modelling framework for UDS, focused on data-based models. The methodology defines a guide to generate data-driven expressions that represent the complex behaviour of network elements using input-output data, expert knowledge and data-fitting. Moreover, the derivation and application of flow-setpoint conversion functions for actuators are presented.

The methodology is assessed in a realistic pilot from Madrid, Spain. A control-oriented model and conversion functions are derived, evaluating them using a real-world rain scenario over virtual reality. The results show high accuracy of the data-based models, as well as significant improvements in terms of discharge reduction and WWTP usage by using these models and functions in MPC.

In future work, we plan to explore state-of-the-art modelling techniques, such as Dynamic Mode Decomposition (DMD), to retrieve linear models from input-output datasets, reducing model complexity and dependence on expert knowledge, yielding convex MPC formulations, increasing computational efficiency. 

% \begin{ack}
% Place acknowledgments here.
% \end{ack}

\section*{DECLARATION OF GENERATIVE AI AND AI-ASSISTED TECHNOLOGIES IN THE WRITING PROCESS}
During the preparation of this work the author(s) used ChatGPT in order to receive wording suggestions. After using this tool/service, the author(s) reviewed and edited the content as needed and take(s) full responsibility for the content of the publication.

\bibliography{ifacconf}             % bib file to produce the bibliography
                                                     % with bibtex (preferred)
                                                   
%\begin{thebibliography}{xx}  % you can also add the bibliography by hand

%\bibitem[Able(1956)]{Abl:56}
%B.C. Able.
%\newblock Nucleic acid content of microscope.
%\newblock \emph{Nature}, 135:\penalty0 7--9, 1956.

%\bibitem[Able et~al.(1954)Able, Tagg, and Rush]{AbTaRu:54}
%B.C. Able, R.A. Tagg, and M.~Rush.
%\newblock Enzyme-catalyzed cellular transanimations.
%\newblock In A.F. Round, editor, \emph{Advances in Enzymology}, volume~2, pages
%  125--247. Academic Press, New York, 3rd edition, 1954.

%\bibitem[Keohane(1958)]{Keo:58}
%\newblock \emph{Power and Interdependence: World Politics in Transitions}.
%\newblock Little, Brown \& Co., Boston, 1958.

%\bibitem[Powers(1985)]{Pow:85}
%\newblock Is there a way out?
%\newblock \emph{Harpers}, pages 35--47, June 1985.

%\bibitem[Soukhanov(1992)]{Heritage:92}
%\newblock \emph{{The American Heritage. Dictionary of the American Language}}.
%\newblock Houghton Mifflin Company, 1992.

%\end{thebibliography}

% \appendix
% \section{A summary of Latin grammar}    % Each appendix must have a short title.
% \section{Some Latin vocabulary}              % Sections and subsections are supported  
                                                                         % in the appendices.
\end{document}